\documentclass[aps,prd,twocolumn,showpacs,superscriptaddress]{revtex4}
\usepackage{graphics}
\usepackage{graphicx}
\usepackage{pstricks}
\usepackage{epsfig}

\begin{document}
\title{Lepton masses and mixing without Yukawa hierarchies}

\author{William A. Ponce}
\affiliation{Instituto de F\'\i sica, Universidad de Antioquia,
A.A. 1226, Medell\'\i n, Colombia.}
\author{Oscar Zapata} 
\affiliation{Instituto de F\'\i sica, Universidad de Antioquia,
A.A. 1226, Medell\'\i n, Colombia.}

\begin{abstract}
{We investigate the neutrino masses and mixing pattern in a version of the $SU(3)_c\otimes SU(3)_L\otimes U(1)_X$ model with one extra exotic charged lepton per family as introduced by Ozer. It is shown that an extended scalar sector, together with a discrete $Z_2$ symmetry,  is able to reproduce a consistent lepton mass spectrum without a hierarchy in the Yukawa coupling constants, the former as a consequence of a carefull balance between one universal see-saw and two radiative mechanisms.}
\end{abstract}


\maketitle

\section{\label{sec:sec1}Introduction}
Even if the Standard Model (SM) has an impressive success, it fails 
to provide an explanation for the fermion masses and mixing angles,
both in the quark and in the lepton sectors. Moreover, recent
experimental results\cite{ossc} confirm that neutrinos have small masses and oscillate.

The solar and atmospheric neutrino oscillations are now well established, and the $\Delta m^2$ values and mixing angles have been measured to the values~\cite{moha}
\begin{eqnarray}\label{expn}\nonumber
\Delta m^2_{atm}&=&2.4(1^{+0.21}_{-0.26})\times 10^{-3}\mbox{eV}^2,\\ \nonumber
\Delta m^2_{sol}&=&7.92(1\pm 0.09)\times 10^{-5}\mbox{eV}^2, \\ \nonumber
\sin^2\theta_{atm}&=&0.44(1^{+0.44}_{-0.22}),\\ \nonumber
\sin^2\theta_{sol}&=&0.314(1^{+0.18}_{-0.15}),\\ 
\sin^2\theta_{chooz}&\leq &0.009 \hspace{.2cm} ,
\end{eqnarray}
which implies, among other things, that at least two neutrinos have very small but non zero masses.

Masses for neutrinos require physics beyond the SM connected either to the existence of right handed neutrinos and/or to the breaking of the B$-$L (baryon minus lepton number) symmetry. If right handed neutrinos exist, the Yukawa terms leads, after electroweak symmetry breaking, to Dirac neutrino masses, requiring Yukawa coupling constants for neutrinos $h_\nu^\phi\leq 10^{-13}$. But the right handed neutrinos, singlets under the SM gauge group, can acquire large Majorana masses and turn on the see-saw mechanism~\cite{seesaw}, an appealing and natural scenery for neutrino mass generation.

The left handed neutrinos, members of the SM lepton doublet $\psi_{lL}^T=(\nu_l,l^-)_L, \; l=e,\mu,\tau $, can also acquire a Majorana mass $m_\nu$ which carries weak isospin I=1 and violates lepton number L by two units, generated either via non renormalizable operators obtained by using twice the SM Higgs doublet $\phi$, or due to coupling with a Higgs triplet $\Delta$ which develops non zero Vacuum Expectation Values (VEV), breaking in this way the lepton number spontaneously which in turn implies the existence of a Majoron~\cite{gelron}.

The alternative Zee mechanism~\cite{zee1} can be implemented, when the L=2 Lorentz scalar $\psi_{lL}C\psi_{l^\prime L}$ (with $C$ the charge conjugation matrix) is coupled to an $SU(2)_L$ charged singlet $h^+$ with L$=-2$, introducing next a new scalar doublet $\phi^\prime$ and breaking the L symmetry explicitly in the scalar potential with a term of the form $\phi\phi^\prime h^+$. In this way neutrino Majorana masses are generated by one loop quantum effects and the Majoron is not present.

The second Higgs doublet $\phi^\prime$ can be avoided by introducing instead a double charged Higgs singlet $k^{++}$ which couples to the single charged one by the trilinear coupling $k^{++}h^-h^-$ and to the right handed charged leptons singlets $l_R^-$ via a term of the form $l_R^-Cl^{\prime -}_Rk^{++}$, generating in this way Majorana small masses via two loop quantum effects by what is known as the Zee-Babu mechanism~\cite{zee2, babu}.

In this paper we explore a model based on the local gauge group $SU(3)_c\otimes SU(3)_L\otimes U(1)_X$~\cite{pf, vl, ozer, sher} (named hereafter 3-3-1 for short), which provides a special framework for Majorana neutrino masses. The original model, which avoids the presence of exotic electric charges but includes heavy electrons~\cite{ozer}, is enlarged with a convenient set of Higgs fields in order to circumvent hierarchies in the Yukawa coupling constants. Besides, the presence of a large scale $V$ related to the breaking of $SU(3)_L\otimes U(1)_X$, triggers a see-saw mechanism for the charged leptons, that provides a realistic mass spectrum for the particles. At the same time, this mechanisms gives relationships that allow one to connect the mass eingenstates with the weak interaction states.

The analysis is done in a similar way that the one presented in Ref.~\cite{canal}, where a related calculation was carried through for the quarks in the 3-3-1 model with right handed neutrinos~\cite{vl}, model that by the way does not contain exotic electrons like the model analyzed here, becoming thus unable to generate see-saw masses for charged leptons, without introducing by hand singlet charged vectorlike exotic electrons.

This paper is organized as follows: In Sec.~\ref{sec:sec2} we review the model with exotic electrons, In Sec.~\ref{sec:sec3} we study the mass spectrum for the charged leptons, in Sec.~\ref{sec:sec4} we study the mass spectrum for the neutrinos. The neutral lepton phenomenology is presented in Sec.~\ref{sec:sec5}, while the final section is devoted to the conclusions.

\section{\label{sec:sec2}The model} 
The model, based on the 3-3-1 local gauge group has 17
gauge Bosons: one gauge field $B_\mu$ associated with $U(1)_X$, the 8
gluon fields $G_\mu$ associated with $SU(3)_c$ which remain massless after
breaking the electroweak symmetry, and 8 gauge fields associated with
$SU(3)_L$ that we write conveniently as \cite{pgs}

\[\sum_{\alpha=1}^8\lambda_\alpha A_\mu^\alpha=\sqrt{2}\left(
\begin{array}{ccc}D_\mu^1 & W^{+}_\mu & K^{+}_\mu \\ W^{-}_\mu & D_\mu^2 &
K^{0}_\mu \\ K^{-}_\mu & \bar{K}^{0}_\mu & D_\mu^3 \end{array}\right), \]
where $D_\mu^1=A^3_\mu/\sqrt{2}+A^8_\mu/\sqrt{6},\;
D_\mu^2=-A^3_\mu/\sqrt{2}+A^8_\mu/\sqrt{6}$, and
$D_\mu^3=-2A^8_\mu/\sqrt{6}$. $\lambda_\alpha, \; \alpha=1,2,...,8$, are the eight Gell-Mann matrices normalized as 
$Tr(\lambda_\alpha\lambda_\beta) =2\delta_{\alpha\beta}$.

The charge operator associated with the unbroken gauge symmetry $U(1)_Q$ 
is 
\begin{equation}
Q=\frac{\lambda_{3L}}{2}+\frac{\lambda_{8L}}{2\sqrt{3}}+XI_3,
\end{equation}
where $I_3=Diag.(1,1,1)$ is the diagonal $3\times 3$ unit matrix, and the 
$X$ values are related to the $U(1)_X$ hypercharge and are fixed by 
anomaly cancellation. 
The sine square of the electroweak mixing angle is given by 
\begin{equation}
S_W^2=3g_1^2/(3g_3^2+4g_1^2),
\end{equation}
where $g_1$ and $g_3$ are the coupling 
constants of $U(1)_X$ and $SU(3)_L$ respectively. 

In the context of this model the quark content for the three families is 
\cite{ozer}: 
$Q^i_{L}=(d^i,u_i,U^i)_L\sim(3,3^*,1/3),\;i=1,2$ for two families,
where $U^i_L$ are two extra quarks of electric charge $2/3$ (the numbers
inside the parenthesis stand for the $[SU(3)_c,SU(3)_L,U(1)_X]$ irreducible representations in that order); $Q^3_{L}=(u^3,d^3,D)_L\sim (3,3,0)$, where
$D_L$ is an extra quark of electric charge $-1/3$. The right handed quarks
are $u^{ac}_{L}\sim (3^*,1,-2/3),\; d^{ac}_{L}\sim (3^*,1,1/3)$ with
$a=1,2,3,$ a family index, $D^{c}_{L}\sim (3^*,1,1/3)$, and
$U^{ic}_L\sim (3^*,1,-2/3)$ for $i=1,2$.

The lepton content of the model, which contribute to the cancellation of 
the anomalies, is given by the following three $SU(3)_L$ triplets: 
$L_{lL} = (\nu_l^0,l^-.E_l^-)_L\sim (1,3,-2/3)$, for 
$l=e,\mu,\tau$, and the six singlets $l^+_{L}, E_{lL}^+\sim(1,1,1)$. Notice 
the presense of three exotic charged leptons $E_l^-$ (one per family), the 
fact that the model does not include right-handed neutrinos, and that 
universality is present at the tree-level in the weak basis.

With the former quantum numbers it is just a matter of counting to check
that the model is free of the following gauge anomalies\cite{pgs}:  
$[SU(3)_c]^3$; (as in the SM $SU(3)_c$ is vectorlike); $[SU(3)_L]^3$ (six
triplets and six anti-triplets), $[SU(3)_c]^2U(1)_X; \; [SU(3)_L]^2U(1)_X
; \;[grav]^2U(1)_X$ and $[U(1)_X]^3$, where $[grav]^2U(1)_X$ stands for
the gravitational anomaly\cite{del}.

\section{\label{sec:sec3}The Charged Lepton Sector}
To the original set of three scalar fields and Vacuum Expectation Values (VEV) introduced in the original paper~\cite{ozer}
\begin{eqnarray*}\label{higgsses}
\langle\phi_1\rangle^T &=&\langle(\phi^+_1, \phi^0_1,\phi^{'0}_1)\rangle = 
\langle(0,v_1,0)\rangle \sim (1,3,1/3) \\ 
\langle\phi_2\rangle^T &=&\langle(\phi^0_2, \phi^-_2,\phi^{'-}_2)\rangle = 
\langle(v_2,0,0)\rangle \sim (1,3,-2/3) \\ 
\langle\phi_3\rangle^T &=&\langle(\phi^+_3, \phi^0_3,\phi^{'0}_3)\rangle = 
\langle(0,0,V)\rangle \sim (1,3,1/3),\\ 
\end{eqnarray*}
let us add a fourth scalar triplet 
\[\langle\phi_4\rangle^T =\langle(\phi^+_4, \phi^0_4,\phi^{'0}_4)\rangle = \langle(0,0,v_4)\rangle \sim (1,3,1/3),\] 
needed in order to implement the universal see-saw mechanism\cite{useesaw} for the charged lepton sector, which also requires of the following discrete $Z_2$ symmetry
\[Z_2(\phi_1,\phi_2,\phi_3,E^+_{lL})=1, \;\;Z_2(\phi_4,L_{lL},l^+_L)=0.\]

There are for this model two mass scales: $v_1\sim v_2\sim v_4\sim 10^2$ GeV the electroweak scale, and $V$ of the order of a few TeV (the 3-3-1 mass scale). Using them, and assuming $v_1=v_2=v_3\equiv v$, we can define an expansion parameter $\delta=v/V$ which triggers the universal see-saw mechanism~\cite{useesaw} for the charged leptons in the context of this model.

The analysis shows that the former set of VEV breaks the 
$SU(3)_c\otimes SU(3)_L\otimes U(1)_X$ symmetry in two steps 
following the scheme
\[\mbox{3-3-1}\stackrel{V}{\longrightarrow}SU(3)_c\otimes SU(2)_L\otimes 
U(1)_Y\stackrel{v}{\longrightarrow} SU(3)_c\otimes U(1)_Q,\] 
which in turn allows for the matching conditions $g_2=g_3$ and 
\[\frac{1}{g^{\prime 2}}=\frac{1}{g_1^2}+\frac{1}{3g_2^2},\]
where $g_2$ and $g^\prime$ are the gauge coupling constants of 
the $SU(2)_L$ and $U(1)_Y$ gauge groups in the SM, respectively.
Notice that only $\phi_1$ and $\phi_2$ contribute to the $W^\pm$ mass, giving a value $M_W^2=g_3(v_1^2+v_2^2)/2=g_2v^2$. Using $M_W=80.425$~\cite{pdb} it implies $v\approx 175/\sqrt{2}\approx 124$ GeV.

The most general Yukawa terms for the charged lepton sector, without making use of the $Z_2$ symmetry, are 
\begin{equation} \label{mlep}
{\cal L}_Y^l=\sum_{\alpha = 1,3,4}\sum_{l,l^\prime = e,\mu,\tau} 
L_{lL}\phi^*_\alpha C(h_{ll^\prime}^{E\alpha}E_{l^\prime}^+
+h_{ll^\prime}^{e\alpha}l^{\prime +})_L + h.c., 
\end{equation}
where $C$ is the charge conjugation matrix. 

In the flavor basis $\vec{E}_6=(e,\; \mu, \; \tau, \; E_e, \; E_\mu, \; E_\tau)$ 
and with the discrete symmetry $Z_2$ enforced, ${\cal L}$ produces the 
following $6\times 6$ mass matrix 
\begin{equation}\label{malep}
M^e=v\left(\begin{array}{cccccc}
0 & 0 & 0 & h^{E1}_{ee} & h^{E1}_{e\mu} & h^{E1}_{e\tau} \\
0 & 0 & 0 & h^{E1}_{\mu e} & h^{E1}_{\mu\mu} & h^{E1}_{\mu\tau} \\
0 & 0 & 0 & h^{E1}_{\tau e} & h^{E1}_{\tau\mu} & h^{E1}_{\tau\tau} \\
h^{e4}_{ee} & h^{e4}_{e\mu} & h^{e4}_{e\tau} & h^{E3}_{ee}\delta^{-1} & h^{E3}_{3e\mu}\delta^{-1}  & h^{E3}_{e\tau}\delta^{-1} \\
h^{e4}_{\mu e} & h^{e4}_{\mu\mu} & h^{e4}_{\mu\tau} & 
h^{E3}_{\mu e}\delta^{-1} & h^{E3}_{\mu\mu}\delta^{-1} & h^{E3}_{\mu\tau}\delta^{-1} \\
h^{e4}_{\tau e} & h^{e4}_{\tau\mu} & h^{e4}_{\tau\tau} &h^{E3}_{\tau e}\delta^{-1} & h^{E3}_{\tau\mu}\delta^{-1}  & 
h^{E3}_{\tau\tau}\delta^{-1} \\
\end{array}\right)
\end{equation}
Assuming for simplicity conservation of the family lepton number in the 
exotic sector ($h^{E3}_{ll^\prime}=h_l\delta_{ll^\prime}$ which does not affect at all the main results), the matrix (\ref{malep}) still remains with 21 Yukawa coupling constants and it is full of physical possibilities. For example, if all the 21 Yukawa coupling constants are different to each 
other (but of the same order of magnitude), we have that $M_eM_e^\dagger$ is a rank six mass matrix, with three eigenvalues of order $V^2$ and three see-saw eigenvalues of order $v^2\delta^2$.

To start the analysis let us impose the permutation symmetry $e\leftrightarrow\mu\leftrightarrow\tau$, make all the Yukawa coupling constants equal to a common value $h_\tau$ and use conservation of the family lepton number in the exotic sector. With these assumptions the following symmetric mass matrix is obtained:

\begin{equation}\label{maalep}
M_0^e=h_\tau v\left(\begin{array}{cc}
M^{ee}_0 & M^{eE}_0 \\
M^{Ee}_0 & M^{EE}_0 \end{array}\right),
\end{equation}
where $M^{ee}_0$ is the $3\times 3$ zero matrix (all the entries equal to zero), $M^{EE}_0=\delta^{-1}I_3$ where $I_3=Diag.(1,1,1)$ as above, and
\begin{equation}
M^{eE}_0=M^{Ee}_0=\left(\begin{array}{ccc}
1&1&1\\
1&1&1\\
1&1&1\\
\end{array}\right),
\end{equation}
are democratic type $3\times 3$ matrices. The subscript zero in the former  matrices means that all the Yukawa coupling constants have been taken equal to a common value, $h_\tau$.

$M_0^e$ in (\ref{maalep}) is a symmetric, rank four see-saw type mass matrix, with the following set of eigenvalues 
\begin{equation}\label{maasee}
h_\tau v[0,0,(\delta^{-1}\pm\sqrt{36+\delta^{-2}})/2,V,V],
\end{equation}
with the two zero eigenvalues related to the null subspace 
$(1,-1,0,0,0,0)/\sqrt{2}$ and $(1,1,-2,0,0,0)/\sqrt{6}$ that we identify in first approximation with the electron and the muon states. Equations (\ref{maalep}) and (\ref{maasee}) implies that the $\tau$ lepton may be identify approximately with the eigenvector $(1,1,1,-3\delta, -3\delta, -3\delta)/\sqrt{3+27\delta^2}]$, with a mass eigenvalue $-m_\tau\approx 9h_\tau v\delta$ (the negative mass value can be fixed either by a change of phase, or either by the transformation $\psi\longrightarrow\gamma_5\psi$ in the Weyl spinor $\psi_R$ of the $\tau$ lepton).

The next step is to break the $e\leftrightarrow\mu\leftrightarrow\tau$ symmetry but just in the $\tau$ sector, keeping for a while the $e\leftrightarrow\mu$ symmetry. This is simple done by letting $h^{E2}_{\tau\tau}=h^{e1}_{\tau\tau}\equiv h_\mu$ of the order of $h_\tau$, with all the other Yukawa coupling consants as in Eq.~(\ref{maalep}). We thus get a rank five symmetric mass matrix, with two see-saw eigenvalues, and a zero mass eigenstate related to the eigenvector $(1,-1,0,0,0,0)/\sqrt{2}$ that we identify with the electron state. The two see-saw eigenvalues, neglecting terms of ${\cal O}(\delta^2)$, are given by
\[\frac{-h_\tau v\delta}{2}[8+(\frac{h_\mu}{h_\tau})^2\pm(2+\frac{h_\mu}{h_\tau})\sqrt{12-4(h_\mu/h_\tau)+(h_\mu/h_\tau)^2}].\]
Using for $m_\tau \approx 1776.99$ Mev and $m_\mu\approx 105.66$ Mev \cite{pdb} it is found that $h_\mu$ must be tuned to the value $h_\mu \approx 2.87 h_\tau$, which in turn implies $m_\tau \approx -15.3 h_\tau v\delta$, value this last one used to obtain the bounds $0.1\leq h_\tau\leq 0.5$, for 13 TeV$\leq V\leq$ 66 TeV.

\begin{figure}
\includegraphics{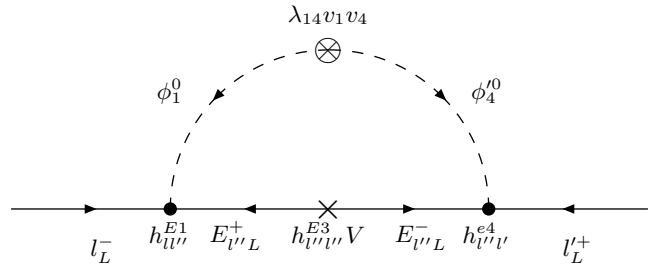}
\caption{\label{fig3}One loop diagram contributing to the radiative 
generation of the electron mass.}
\end{figure}

The consistency of the model requires now to find a mechanism able to generate an appropriate mass for the electron, avoiding as much as possible a hierarchy in the Yukawa coupling constants ($M^e$ in Eq.~(\ref{malep}) by itself is able to generate such a mass at tree level, but at the expense of an unwanted hierarchy). For this purpose the radiative mechanism~\cite{babma} can be implemented by using the rich scalar sector of the model. As a matter of fact, forbiding triple Higgs scalars couplings in the scalar potential, the three diagrams in Fig.~\ref{fig3} can be extracted from the Lagrangean (one for each exotic charged lepton), where the mixing in the Higgs sector is coming from a term in the scalar potential of the form $\lambda_{14}(\phi_1^*.\phi_1)(\phi_4^*.\phi_4)$.

The contribution given by this diagram, assuming validity of the Extended Survival Hypothesis (ESH)~\cite{esh} that for this situation implies  $M_{\phi_1^0}\approx M_{\phi_4^{\prime 0}}\approx v$, is 
\begin{equation}\label{mlee}
(M_1^{ee})_{ll^\prime}= \frac{\lambda_{14}h_\tau^2\delta\ln\delta}
{8\pi^2}(M_1^{eE}.M_1^{Ee})_{ll^\prime},
\end{equation}
where 
\begin{equation}\label{mlEe}
M_1^{eE}=M_1^{Ee}=\left(\begin{array}{ccc}
1& 1 & 1 \\
1 & 1 & 1 \\
1 & 1 & h_\mu/h_\tau \\
\end{array}\right),
\end{equation}
are the new upper right, and lower left $3\times 3$ corners respectively  in matrix $M_1^e$, with the subscript 1 meaning that the Yukawa coupling constants are taken such that they are suited to reproduce the masses for the $\mu$ and the $\tau$ leptons ($M_1^{EE}=M_0^{EE}$).

Clearly, $M_1^{ee}$ in Eq.~(\ref{mlee}) is a democratic mass submatrix in the $2\times 2$ upper left corner, which means that it is unable to generate a mass for the electron. The alternative at hand is to slightly  break the $e\leftrightarrow\mu$ symmetry present in the mass matrix (\ref{maalep}). This is achieved by letting $h_{ee}^{E1}=h_\tau(1-k_e/h_\tau)$ and $h_{ee}^{e4}=h_\tau(1+k_e/h_\tau)$, with $k_e\sim 10^{-1}$, a parameter related to the electron mass ($m_e=0$ for $k_e=0)$. Using this final set of Yukawa coupling constants we get 

\[m_e= -\frac{\lambda_{14} k_e^2 v\delta\ln\delta }{4\pi^2},\]
which for $m_e=0.51$ MeV~\cite{pdb}, $\lambda_{14}=0.5$, $V= 13$ TeV (implying $h_\tau =0.1$), and $v=124$ GeV, produces the value $k_e\approx 0.08$, in agreement with our original assumption.

In this way we have arrived to the final form of the charged lepton mass matrix 
\begin{equation}\label{mppe}
M_2^e=h_\tau v\left(\begin{array}{cc}
M^{ee}_2 & M^{eE}_2 \\
M^{Ee}_2 & M^{EE}_2 \end{array}\right),
\end{equation}
where $M_2^{ee}=M_1^{ee}$ as given by Eq.(\ref{mlee}), $M_2^{EE}=M_0^{EE}=\delta^{-1}I_3$, 
\begin{equation}
M_2^{eE}=\left(\begin{array}{ccc}
1-k_e/h_\tau & 1 & 1 \\
1 & 1 & 1 \\
1 & 1 & h_\mu/h_\tau \\
\end{array}\right),
\end{equation}
and
\begin{equation}
M_2^{Ee}=\left(\begin{array}{ccc}
1+k_e/h_\tau & 1 & 1 \\
1 & 1 & 1 \\
1 & 1 & h_\mu/h_\tau \\
\end{array}\right).
\end{equation}
Notice that with the tuning $h_\mu\approx 2.87h_\tau$ and $k_e\approx 0.08, \; M_2^e$ in Eq.~(\ref{mppe}) reproduces the charged lepton mass spectrum without a hierarchy in the Yukawa coupling constants. as far as we include the one loop quantum effects (if they are not included, $K_e\approx 0.08$ by itself, gives a too small electron mass).

\section{\label{sec:sec4}The Neutral Lepton Sector}
With the particle content introduced so far there are not tree-level mass terms for the neutrinos, neither radiative corrections able to produce them. Masses for the neutral lepton sector are obtained only by enlarging the model with extra fields.

Radiative Majorana masses for the neutrinos can be generated when a new scalar triplet 
$\phi_5=(\phi_5^{++}, \phi_5^+ ,\phi_5^{\prime +})\sim (1,3,4/3)$ is introduced, with a $Z_2$ charge 
equal to zero (notice that $\langle\phi_5\rangle\equiv 0$). This new scalar triplet couple to the spin 1/2 leptons via a term in the Lagrangian of the form:
\begin{eqnarray}\label{massn}\nonumber
{\cal L} &=& \sum_{ll^\prime}h^\nu_{ll^\prime}L_{lL}L_{l^\prime L}\phi_5 = 
\sum_{ll^\prime}h^\nu_{ll^\prime}[\phi_5^{++}(l^-_LE^-_{l^\prime L}-l^{\prime -}_LE^-_{lL}) \\ 
&+& \phi_5^{+}(E^-_{lL}\nu_{l^\prime L}-E^-_{l^\prime L}\nu_{lL})
+\phi_5^{\prime +}(\nu_{lL}l^{\prime -}_L-\nu_{l^{\prime}L}l^-_L)],
\end{eqnarray}
for $l\neq l^\prime = e,\mu,\tau$. The three new parameters $h^\nu_{ll^\prime}$ introduced are going to be used in Sec.\ref{sec:sec5} for the phenomenology of the neutrino masses and mixing angles. 

Contrary to other 3-3-1 models where the lepton number L does not commute  with the gauge group~\cite{liu}, for this particular model L is a good quantum number as can be seen from the following lepton number assignment for the multiplets: L$(L_{lL})=-$L$(l^+_L,E^+_L)=1$ for $l=e,\mu,\tau$, and L=0 for all the other multiplets, including gauge Boson fields and also for the original four scalar Higgs field triplets $\phi_\alpha, \; \alpha=1,2,3,4$. Then, conservation of the lepton number L in Eq.(\ref{massn}) requires that L$(\phi_5)=-2$. With this assignment of L, which by the way reproduces the assignment of the lepton numbers for the multiplets of the SM, L can not be broken spontaneously. It can be broken only explicit in the scalar potential which means that no Majoron is going to be present in the context of this model. As a matter of fact, a term in the scalar potential of the form $\lambda_{34}(\phi_3^*.\phi_5)(\phi_4^*.\phi_2)$ which violates lepton number by two units, turns on the radiative mechanism~\cite{zee1, babma} for generating neutrino masses and mixing in the context of this model.

\begin{figure}
\includegraphics{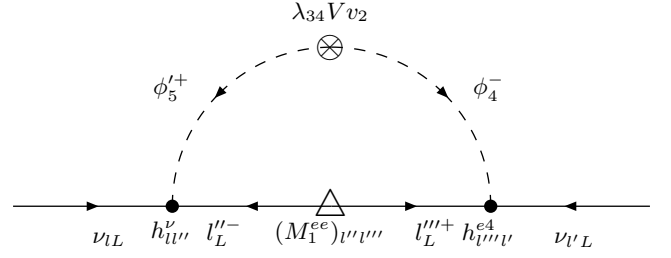}
\caption{\label{fig4}Loop diagrams contributing to the radiative 
generation of Majorana masses for the neutrinos.}
\end{figure}

The previous ingredients allow us to draw the diagram in Fig.~\ref{fig4}, a second order radiative diagram (the mass insertion $M_1^{ee}$ corresponds to the diagram in Fig~\ref{fig3}) which produces the following neutrinos mass matrix in the flavor basis $\vec{\nu}_{Lf}^T=(\nu_e,\nu_\mu,\nu_\tau)_L$ 
\begin{equation}\label{mnu}
(M_\nu)_{ll^\prime}= \frac{\lambda_{34}h_{\tau}^2\delta\ln{\delta}}{8\pi^2}
(H.M^{ee}_1.M_2^{eE})_{ll^\prime}, 
\end{equation}
where the $3\times 3$ matrices $M_1^{ee}$ and $M_2^{eE}$ in their analytical form were presented in the previous section and 
\begin{equation}\label{mah}
H=\left(\begin{array}{ccc}
0 & h^\nu_{e\mu} & h^\nu_{e\tau} \\
-h^\nu_{e\mu} & 0 & h^\nu_{\mu\tau} \\
-h^\nu_{e\tau} & -h^\nu_{\mu\tau} & 0 
\end{array} \right),
\end{equation}
with the Yukawa coupling constants $h^\nu_{ll^\prime}$ taken as free parameters that are going to be fixed anon.

\section{\label{sec:sec5}Phenomenology}
The lepton mass terms in the Lagrangean can be written in the context of this model as
\begin{equation}\label{nmass}
{\cal L}_m=\vec{\nu}_{Lf}^TM^{\prime\nu} C\vec{\nu}_{Lf}+\vec{E}_{6L}^TM_e^{\prime\prime}C\vec{E}^c_{6L} + h.c., 
\end{equation}
where $\vec{\nu}_{Lf}^T$ and $\vec{E}_6$ are vectors in the flavor basis as defined before, $M^{\prime}_\nu =(M_\nu+M^{\nu}_T)/2$ is the symmetric $3\times 3$ neutrino mass matrix constructed from $M_\nu$ in Eq.~(\ref{mnu}) and $M_e^{\prime\prime}$ is the $6\times 6$ charged lepton mass matrix in Eq.~(\ref{mppe}).

Relations between the flavor states and the mass eigenstates $\vec{\nu}_L=(\nu_1,\nu_2,\nu_3)_L$ are thus provided by the relationship $\vec{\nu}_{Lf}=U_{PMNS}\vec{\nu}_L$, where $U_{PMNS}$ is the $3\times 3$ Pontecorvo - Maki - Nakagawa - Sakata lepton mixing matrix~\cite{ponte}.

The former mass matrices are diagonalize in the following way
\[U_\nu^T M^\prime_\nu U_\nu=M_\nu^d,\;\;\;\;\; U_lM^{\prime\prime}_eV_l^\dagger=M_e^d ,\]
where $M_e^d=Diag.(m_e,m_\mu,m_\tau,M_{E_e},M_{E_\mu},M_{E_\tau})$ and 
$M_\nu^d=Diag.(m_1,m_2,m_3)$ 
are the diagonal mass matrices in the charged and neutral lepton sectors respectively. $U_\nu$ is a $3\times 3$ rotation matrix and $U_l$ and $V_l$ are two $6\times 6$ rotation matrices, the first one transforming the left-handed states and the second one transforming the right handed states and irrelevant to the analysis presented here.

The three matrices $U_\nu, \; U_l$ and $V_l$ connect the flavor states with the mass eigenstates. The physical neutrino mixing matrix is then given, in the context of this model, by 
\begin{equation}
U_{PMNS}=U^\dagger_{3l}U_\nu, 
\end{equation}
where $U_{3l}$ is the $3\times 3$ upper left submatrix of $U_l$. As a consequence of this, the physical neutrino mixing matrix in the context of this model differs from unitarity by terms proportional to $\delta^2\sim 10^{-4}$. At this point we proposse to diagonalize $M_e^{\prime\prime}$ in Eq.~(\ref{mppe}) doing a perturbative calculation, taking as the zero order the matrix $M_e^\prime$ in Eq.(\ref{maalep}) which implies
\begin{equation}
U_{3l}^0=\left(\begin{array}{ccc}
1/\sqrt{2} & -1/\sqrt{2} & 0 \\
1/\sqrt{6} & 1/\sqrt{6} & -2/\sqrt{6} \\
1/\sqrt{3}& 1/\sqrt{3} & 1/\sqrt{3} \\
\end{array}\right),
\end{equation}
which is unitary.

In the literature the unitary mixing matrix for Majorana neutrinos is convenient parameterized as 
\[U_{PMNS}^0 =U_{23}(\theta_{23})U_{13}(\theta_{13},\delta)U_{12}(\theta_{12}) I_\phi,\]
where $U_{ij}$ are rotation matrices in the $ij$ plane by the angle $\theta_{ij}$ and $\delta$ and $I_\phi$ are Dirac and Majorana CP violating phases which are not relevant for our study. This parameterization allows to connect immediately the rotation angles with physical observable because, in first approximation $\theta_{23}=\theta_{atm},\;\theta_{12}=\theta_{sol}$ and $\theta_{13}=\theta_{chooz}$.

The random numerical analysis for $h_{ll^\prime}^\nu$ as the aleatoric variables taken in the range $0.05\leq h_{ll^\prime}^\nu\leq 1.0$, and using Mathematica Monte Carlo subroutines, showed that for $h^\nu_{e\mu}\approx h^\nu_{\mu\tau}\approx h^\nu_{e\tau}\approx 0.1$ it is possible to reproduce the experimental results in  Eq.(\ref{expn})  up to $3\sigma$ deviations and for the inverted hierarchy of the neutrino spectrum~\cite{moha}, with the following predictions for the neutrino masses: $m_{\nu_1}= 0.0491\pm 0.0001$ eV, $m_{\nu_2}= 0.0483\pm 0.0001$ eV, and $m_{\nu_3}= 0.0016278\pm 3\times 10^{-7}$ eV, where the errors are statistical ones.

But $U_{PMNS}$ is not unitary in he context of this model (in general any model with physics beyond the SM produces a neutrino mixing matrix which is not unitary~\cite{zhi}). In order to take into account this fact let us parameterize, in first approximation, such violation with three new parameters $\epsilon_i,\; i=1,2,3$ such that 
\[(U_{PMNS}U_{PMNS}^\dagger)_{ij}\approx \delta_{ij}(1+\epsilon_i)\]

These new parameters $\epsilon_i$ can be evaluated in first order in perturbation theory, keeping terms up to $\delta^2$, by the use of 
$M_e^{\prime\prime}$ in Eq.~(\ref{mppe}) which implies a diagonalization matrix $U_{3l}^\prime \approx U_{3l}^0+\Delta U_{3l}$, where 
\begin{equation}
\Delta U_{3l}=\frac{4.5\delta^2}{\sqrt{3}}\left(\begin{array}{ccc}
0 & 0 & 0 \\
0 & 0 & 0 \\
1 & 1 & 1 \\
\end{array}\right).
\end{equation}
The analysis shows that $\epsilon_1 \approx\epsilon_2\approx 0$ and $\epsilon_3\approx \delta^2\sim 10^{-4}$, a value which implies minor changes to the previous analysis, all of them wash out by the experimental uncertainties in Eq.~(\ref{expn}). So, the violation of unitarity in $U_{PMNS}$ is relatively small in the context of the model studied here (second order corrections are proportional to $\delta^4$).

Since $h^\nu_{ll^\prime}\sim 10^{-1}$, we should assure that the model does not violate the experimental constraints on flavor changing neutral currents (FCNC). In particular, the decay $\mu\longrightarrow e\gamma $ may occur in the context of the model at one loop level due to the coupling of $\phi_5^{\prime +}$ with leptons in the Yukawa Lagrangean in Eq.~(\ref{massn}). The experimental Branching Ratio BR$(\mu\longrightarrow e\gamma)
<1.2 \times 10^{-11}$~\cite{pdb}, which is the strongest constraint in this kind of family lepton number violating processes, gives the restriction~\cite{babu}

\[\frac{h^\nu_{\mu\tau}h^\nu_{\tau e}}{m^2_{\phi^\prime_5}}<0.4\times 10^{-8},\]
which for $m_{\phi^\prime_5}\approx V\sim$ 13 TeV (according to the ESH~\cite{esh}) is satisfied. [2.5BR($\tau\longrightarrow\mu\gamma)\approx$ BR($\tau\longrightarrow e\gamma) <2.7\times 10^{-6}$ are weaker constraints]

The process $\mu^-\longrightarrow e^+e^-e^-$ whose BR is $<1.0\times 10^{-12}$~\cite{pdb} does not occur at tree level in the context of this model because there is not present a double charged Higgs scalar singlet under $SU(2)_L$ as it does for example for the Zee-Babu 
model~\cite{zee2, babu}. In this model the process occurs via box diagrams proportional to $(h_{ll^\prime}^\nu)^4$, highly suppressed by four propagator of heavy particles: two with $\phi_5^{++}$ propagating and the other two with $E_l^-$. The same argument applies to the other FCNC process 
$\tau^-\longrightarrow l^{+}l^{-}l^{\prime}$ (as for example $\tau^-\longrightarrow e^+e^-\mu^-$ etc.) with BR $<1.5\times 10^{-6}$~\cite{pdb}.

Finally notice two things: first, the Higgs scalar $\phi_2$ introduced in the original paper, is not needed for the analysis presented here in the lepton sector (neither it is needed to properly break the symmetry, but it is a fundamental piece in order to provide a consistent quark mass spectrum~\cite{ozer}). Second, $\phi_5$ does not couple to the quark sector of the model; because of this the model does not present the potential threaten of a possible tree level neutrinoless double beta decay.

\section{\label{sec:sec6}Summary and Conclusions}
The main motivation of this study was to investigate the neutrino mass spectrum and mixing pattern in the framework of a model based on the  $SU(3)_c\otimes SU(3)_L\otimes U(1)_X$ local gauge group with exotic electrons, a natural scenario for implementing the universal see-saw mechanism in the charged lepton sector, something we achieved by enlarging the Higgs scalar sector and the introduction of a $Z_2$ discrete symmetry.

The lepton number L in the context of this model is a well defined quantum number and it is violated only explicitly in the scalar potential. As a consequence, the radiative mechanism for generating neutrino masses appears and no Majoron is present.

The analysis was made trying to avoid hierarchies in the Yukawa coupling constants. In the charged lepton sector the hierarchies were avoided by a carefull balance between a see-saw and a radiative mechanisms. The see-saw mechanism is not present for the neutrinos, and the mass difference between neutral and charged leptons is a natural consequence of a two loop versus one loop quantum origen. All the above with a good dosage of fine tuning in some of the free parameters of the model.

Notice that the two-loop quantum corrections involved in this analysis, and the origen of the neutrino masses and mixing, are different from the Zee-Babu~\cite{zee2, babu} type mechanism, with all the free parameters of the model well constrained by the current neutrino oscillation data, with bounds that do not violate FCNC constraints, neither neutrinoless double beta decay limits.

The analysis presented here not only accommodates the neutrino mixing and oscillations, but it also throws neat prediction for the neutrino mass spectrum. Besides, the lower bounds on the masses of the exotic Higgs scalars of the model, could be probed at the upcoming Large Hadron Collider.

Neutrino masses and oscillations in the context of the model analyzed here were studied for the first time in Ref.~\cite{kita}. The main difference between that paper and this one is that in Ref.~\cite{kita} it was assumed diagonal tree level masses for the charged leptons and the implementation of the universal see saw mechanism in the charged lepton sector was not even attempted there, ending them up with the known hierarchy in the Yukawa coupling constants of the lepton sector. Other important difference between the two papers is that in Ref.~\cite{kita}, and in order to implement the Zee-Babu mechanism~\cite{zee2, babu} for generating neutrino mass terms, a double charged Higgs scalar $SU(3)_L$ singlet $k^{++}\sim (1,1,2)$ was used instead of our $\phi_4$ scalar triplet, which is the main ingredient in our analysis for implementing the universal see saw mechanism (the other four scalars are the same in the two papers). So, both papers address to the same problem from two different points of view.

The final values obtained for the 24 Yukawa coupling constants (21 in the charged lepton sector and 3 in the neutral one) are quite surprising even for us. Our original goal was to fit the experimental measurements using Yukawa coupling constants between $0.1\leq h_{ll^\prime}\leq 5$. At the end, most of them became (almost) equal to 0.1 and we do not have any explanation for the kind of symmetries responsable for this amusing result.

{\bf Acknowledgments.} We thank Enrico Nardi for a critical reading of the original manuscript.


\begin{thebibliography}{}
\bibitem[1]{ossc}
Y.Fukuda \textit{et al} (SuperKamiokande Collaboration), Phys. Rev. Lett. \textbf{81}, 1562 (1998); Q.R.Ahmad \textit{et al} (SNO Collaboration), Phys. Rev. Lett. \textbf{89}, 011301 (2002); K.Eguchi \textit{et al} (KamLAND Collaboration), Phys. Rev. Lett. \textbf{90}, 021802 (2003); Y.Ashie \textit{et al}, Phys. Rev. Lett. \textbf{93}, 101801 (2004).

\bibitem[2]{moha}
For recent reviews see: R.N.Mohapatra and A.Y.Smirnov, ``\textit{The neutrino mass and new physics}", [hep-ph/0603118], to appear in the Annual Review of Nuclear and Particle Science, Vol.~\textbf{56}; G.Altarelli, Nucl. Phys. B, Proc. Suppl.~\textbf{143}, 470 (2005).

\bibitem[3]{seesaw}
P.Minkowski, Phys. Lett. B~\textbf{67}, 421 (1977); 
M.Gell-Mann, P.Ramond, and R. Slansky, {\it Supergravity} (P.van Nieuwenhuizen at al eds.), North Holland, Amsterdam, 1980, p.315; T.Yanahida, in {\it Proceedings of the Workshop on the Unified theory and the Baryon number in the Universe} (O.Sawada and A.Sugamoto, eds.), KEK, Tsukuba, Japan (1979), p.95.

\bibitem[4]{gelron}
G.Gelmini and M.Roncadelli, Phys.Lett. B~\textbf{99},411(1981).

\bibitem[5]{zee1}
A.Zee, Phys. Lett. B~\textbf{93}, 389 (1980); \textit{ibid}, B~\textbf{161}, 141 (1985); D.Chang and A.Zee, Phys.Rev.D~\textbf{61}, 071303(R) (2000).

\bibitem[6]{zee2}
A.Zee, Nucl. Phys. B~\textbf{246}, 99 (1986).

\bibitem[7]{babu}
K.S.Babu, Phys. Lett B\textbf{203}, 132 (1988); K.S.Babu and C. Macesanu, Phys. Rev. D~\textbf{67}, 073010 (2003).


\bibitem[8]{pf}
F. Pisano and V. Pleitez, Phys. Rev. D \textbf{46}, 410 (1992);  P.H.
Frampton, Phys. Rev. Lett. \textbf{69}, 2889 (1992); V. Pleitez and M.D. Tonasse, Phys. Rev. D \textbf{48}, 2353 (1993);
\textit{ibid} D\textbf{48}, 5274 (1993);  D. Ng, Phys. Rev. D \textbf{49}, 4805 (1994);
L. Epele, H. Fanchiotti, C. Garc\'\i a Canal and D. G\'omez Dumm, Phys.
Lett. B \textbf{343} 291 (1995);  M. \"Ozer, Phys. Rev. D \textbf{54},
4561 (1996).

\bibitem[9]{vl}
J.C.Montero, F.Pisano and V.Pleitez, Phys. Rev. D\textbf{47}, 2918 (1993);  
R. Foot, H.N. Long and T.A. Tran, Phys. Rev. D \textbf{50}, R34 (1994). 

\bibitem[10]{ozer}
M.\"Ozer, Phys. Rev. D\textbf{54}, 1143 (1996).

\bibitem[11]{sher}
D.L.Anderson and M.Sher, Phys. Rev. D~\textbf{72}, 095014 (2005).

\bibitem[12]{canal}
H.Fanchiotti, C.Garc\'\i a-Canal, and W.A Ponce, Eur. Phys. Lett. \textbf{72}, 733 (2005); D.A.Guti\'errez, W.A.Ponce, and L.A.S\'anchez, Eur. Phys. J. C~\textbf{46}, 497 (2006).

\bibitem[13]{pgs}
W.A. Ponce, Y. Giraldo and L.A. S\'anchez, Phys. Rev. D \textbf{67}, 075001 
(2003).

\bibitem[14]{del}
R.Delbourgo, Phys. Lett. \textbf{40B}, 381 (1972); L.Alvarez-Gaume, Nucl. 
Phys. \textbf{B234}, 262 (1984).

\bibitem[15]{useesaw}
A.Davidson and K.C.Wali, Phys. Rev. Lett., \textbf{59}, 393 (1987); S.Rajpoot, Phys. Rev. D~\textbf{36}, 1479 (1987); D.Chang and R.N.Mohapatra, Phys. Rev. Lett.,~\textbf{58}, 1600 (1987); B.S.Balakrishna, Phys. Rev. Lett. \textbf{60}, 1602 (1988).

\bibitem[16]{pdb}
Particle Data Group, W.-M Yao \textit{et al}, J. Phys. G~\textbf{33}, 1 (2006).

\bibitem[17]{babma}
K.S.Babu and E.Ma, Mod. Phys. Lett. A~\textbf{4}, 1975 (1989).

\bibitem[18]{esh}
F. del Aguila and L. Iba\~nez, Nucl. Phys. B~\textbf{177}, 60 (1981).

\bibitem[19]{liu}
J.T.Liu and D.Ng, Phys. Rev. D~\textbf{50}, 548 (1994); M.B.Tully and G.C.Joshi, Phys. Rev. D~\textbf{64}, 011301(R), (2001); D.Chang and H.N.Long, Phys. Rev. D~\textbf{73}, 053006 (2006).

\bibitem[20]{ponte}
B.Pontecorvo, Zh. Eksp. Theor. Fiz.~\textbf{34} 247 (1958); Z.Maki, M.Nakagawa, and S.Sakata, Phys.~\textbf{28}, 870 (1962).

\bibitem[21]{zhi}
S. Antusch \textit{et al}: ``Unitarity of the Leptonic Mixing Matrix",
[hep-ph/0607020]; Z.-Z. Xing and S. Zhou: ``Why is the $3\times 3$ neutrino mixing matrix almost unitary in realistic seesaw models" [hep-ph/0512290].

\bibitem[22]{kita}
T. Kitabayashi, Phys. Rev. D~\textbf{64}, 057301 (2001).

\end{thebibliography}
\end{document}